\documentclass[aps,prd,showpacs,floatfix,twocolumn,nofootinbib,superscriptaddress]{revtex4-1} %
\usepackage{dcolumn}% Align table columns on decimal point
\usepackage{amsmath,amsfonts,amssymb,mathrsfs,times,bm,color}
\usepackage{extarrows}
\usepackage{graphicx}
\usepackage{float}
\usepackage{booktabs}
\usepackage{graphicx,epstopdf}% Include figure files
\usepackage{dcolumn}% Align table columns on decimal point
\usepackage{exscale}
\usepackage{relsize}
\usepackage{latexsym}
\usepackage{longtable}
\usepackage[colorlinks=true,breaklinks=true,linkcolor=blue,citecolor=blue,urlcolor=blue]{hyperref}

\newcommand{\nn}{\nonumber}

\newcommand{\orcid}[1]{\href{https://orcid.org/#1}{\includegraphics[scale=0.035]{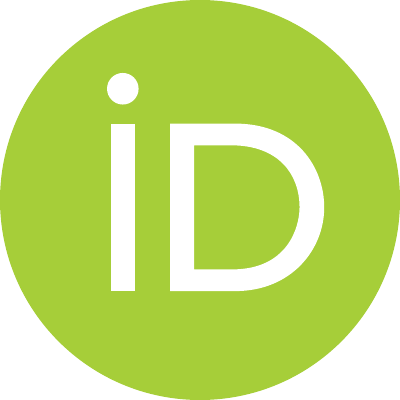}}}

\begin{document}

\title{Leading-order gravitational time delay of massive particles by a moving Schwarzschild lens}

\author{Guansheng He\hspace*{0.6pt}\orcid{0000-0002-6145-0449}\hspace*{0.8pt}}
\affiliation{Purple Mountain Observatory, Chinese Academy of Sciences, Nanjing 210023, China}
\affiliation{School of Mathematics and Physics, University of South China, Hengyang 421001, China}
\author{Wenbin Lin\hspace*{0.6pt}\orcid{0000-0002-4282-066X}\hspace*{0.8pt}}
\affiliation{School of Mathematics and Physics, University of South China, Hengyang 421001, China}
\affiliation{School of Physical Science and Technology, Southwest Jiaotong University, Chengdu 610031, China}
\author{Yi Xie\hspace*{0.6pt}\orcid{0000-0003-3413-7032}\hspace*{0.8pt}}
\email{Corresponding author. yixie@pmo.ac.cn}
\affiliation{Purple Mountain Observatory, Chinese Academy of Sciences, Nanjing 210023, China}

\date{\today}

\begin{abstract}
The leading-order gravitational time delay of relativistic neutral massive particles (e.g., neutrinos or high-energy cosmic-ray particles) caused by a moving Schwarzschild black hole with a constant radial velocity is investigated for the first time. On the basis of the equations of motion in the spacetime of the moving lens, we achieve a new unified formula for the travel times of relativistic massive and massless particles propagating from the source to the observer within the first post-Minkowskian approximation. The analytical form of the difference between the travel times of a relativistic massive particle and a light signal in this geometry, as well as that of two relativistic massive particles, is thus obtained in the weak-field and slow-motion limit. The influence of the radial lens motion on the leading-order Schwarzschild time delay of relativistic massive particles is then discussed. It is found that in the slow-motion limit, the radial lens motion towards the observer decreases the flight time of an ultrarelativistic massive particle, when compared with the case of no translational motion of the central body. Conversely, if the lens gets away from the detector radially under the same conditions, the propagation process of the particle will slow down and its flight time will thus increase in comparison with the Schwarzschild case. Finally, we analyze the magnitude of the full radial motion effect of the lens and evaluate the possibility of its astronomical detection by modeling three typical black holes as the lens respectively.

\end{abstract}

%\pacs{04.20.-q, 95.30.Sf, 98.62.Sb}

\maketitle

\section{Introduction} \label{sect1}

The time dependence of a gravitational field originated from the translational motion of the central body relative to the observer usually exerts an influence on the propagation of light signals or non-photonic messengers, which is the so-called velocity effect~\cite{PB1993,WS2004,Heyrov2005}. Since the pioneering work of Birkinshaw and Gull~\cite{BG1983}, where an ingenious approach for measuring the transverse velocities of clusters of galaxies was proposed, kinematical effect of this kind has attracted more and more attention (see, for instance,
~\cite{BFP1998,KS1999,FKN2002,KP2003,KM2007,KF2007,XH2008,WX2008,DXH2009,DX2012,Xie2014,BZ2015,Deng2015,Zscho2018,Zscho2019,YCC2024}) due to its special significance in modern astrophysics and fundamental astrometry. The main motivations responsible for an increasing interest on this effect lie in three aspects. First, we know that many related observable relativistic effects of test particles may be eventually influenced by the motion of the gravitational system. Providing the gravitational body (e.g., a star or a stellar-mass black hole) moves quickly, the velocity effects will be so noticeable that they may affect the high-accuracy measurements of some gravitational effects~\cite{KS1999,HL2016a}. This would not be impossible, since high-velocity or hypervelocity celestial bodies are not rare in our universe. For example, the transverse velocity of the pulsar B$1508+55$ reaches $1083^{+103}_{-90}\,$km/s~\cite{CVBCCTFLK2005}. Hence, a further theoretical probe of the velocity effects on the leading-order contributions (and even the second- or higher-order contributions) to various gravitational effects for different types of astronomical scenarios and background spacetimes is necessary. A second aspect is that the rapid improvements in instruments and techniques of high-accuracy astronomical observations (see~\cite{Perryman2001,Trippe2010,ZRMZBDX2013,RH2014,RD2020,Brown2021,LXLWBLYHL2022,LXBLLLH2022}, and references therein), along with the remarkable achievements in the multi-messenger synergic observations~\cite{BT2017,IceCube2018,MFHM2019,Huerta2019,Poemma2021,QJFZZ2021}, have been made over the last decades. A representative technique is very long baseline interferometry~\cite{Burke1969,Rogers1970,TMS1986,Hirabayashi1998,Ma1998,Raymond2024}, the next-generation system of which proposes an extremely high precision of 4 picoseconds in measuring the differential time delay of radio signals~\cite{SB2012,Niell2018}. And a time resolution of about 2ns has been achieved by the Large High Altitude Air Shower Observatory (LHAASO) of cosmic-ray particles~\cite{Cao2024,Ma2022} and by the IceCube neutrino observatory~\cite{HK2010,IceCube2021,IceCube2006,Abbasi2010}. This in turn inspires the attempts to discuss the influence of the translational motion of the gravitational source on the relativistic effects of multiple messengers including non-photonic messengers. A final aspect is indicated by the fact that the translational motion of a moving gravitational system (e.g., a black hole or a neutron star) is not uncommon and could be produced via several different avenues. For instance, a black hole, formed from the merger of two unequal-mass black holes, may receive a gravitational kick due to the asymmetric loss of linear momentum in the gravitational radiation to get a recoil velocity~\cite{GSBHH2007,RDS2012,CCLS2018}. It thus makes the theoretical consideration of the effect of the central body's motion on multifarious gravitational effects more realistic and meaningful.

The gravitational time delay (also called the Shapiro time delay~\cite{Shapiro1964,Shapiro1971}) of electromagnetic waves caused by a central body is an important classical relativistic effect, and the velocity effects on the relativistic time delay of lightlike signals have been investigated in detail in the last three decades~\cite{KS1999,Sereno2002,KM2002,Frittelli2003,Sereno2005,KF2007,BAL2008,Kopeikin2009,HBL2014,SH2014}. In particular, Kopeikin and Sch\"{a}fer~\cite{KS1999} constructed a Lorentz covariant theory of light propagation in the weak field of an arbitrarily moving N-body gravitational system, and obtained an analytical expression for the generalized gravitational time delay of photons in this spacetime in the first post-Minkowskian (PM) approximation. Hees \emph{et al.}~\cite{HBL2014} adopted the time transfer functions~\cite{LLT2004,TL2008} to investigate
 the lightlike Shapiro delay effect due to an ensemble of uniformly moving axisymmetric bodies. The effects of the radial motion of a Kerr-Newman black hole on the equatorial gravitational delay of light up to the 2PM order were also studied~\cite{HL2016b}.

However, to our knowledge no effort has yet been made to consider the influence of the motion of a gravitational system on the Shapiro time delay of timelike particles. Actually, with the coming of the era of multi-messenger astronomy and the increase of mono-messenger or joint multi-messenger astronomical observations, it is of significance and of interest to perform a full theoretical analysis of the gravitational time delay of massive particles induced by a moving body. A first reason is that massive messengers, similar to photonic messengers, can also provide specific information related to our physical universe individually or collectively~\cite{MFHM2019,Huerta2019}.
And it is possible to get additional information about the properties of the gravitational body and the particle source or to place supplementary constraints on the spacetime parameters through the probe of the time delay of timelike signals. Second, the deviation of the initial velocity of a massive particle at infinity from the speed of light may make the observable Shapiro delay of the particle flying from the source to the observer in a stationary geometry more noticeable than its optical counterpart under the same conditions~\cite{JL2019,LJ2020b,HL2022,HXJL2024}. This characteristic is important, because it (i) makes the consideration of the second- and even higher-order contributions to the gravitational delay nontrivial, and (ii) might yield a larger opportunity for observing the Shapiro delay effects of the test particles. It should be fair to mention that further efforts on the issue of the velocity effects on the Shapiro delay of timelike particles are worthwhile and necessary.

In present work, we study the gravitational time delay of relativistic massive particles induced by a radially moving Schwarzschild black hole within the framework of the 1PM approximation for the first time. Starting with the equations of motion of test particles in the spacetime of the moving lens, we achieve a new unified formula for the travel times up to the 1PM order of relativistic massive and massless particles, and then discuss the influence of the radial lens motion on the leading-order Schwarzschild delay of relativistic massive particles. The expression of the difference between the travel times of a relativistic massive particle and a light signal in this geometry, along with that of two relativistic massive particles, is thus obtained in the weak-field and slow-motion approximation. Furthermore, we analyze the magnitude of the full radial velocity effect of the lens and estimate the possibility of its astronomical detection in detail, via modeling three typical black holes (i.e., the stellar-mass black hole in binary star system Cygnus X-1, the intermediate-mass black hole in the galaxy NGC 3319, and the Galactic supermassive black hole) as the lens. Our discussions are restricted in weak-field, small-angle, and thin-lens approximation.

This paper is organized as follows. In Section~\ref{sect2}, we review the weak-field metric of a radially moving Schwarzschild black hole and the equations of motion of test particles in this spacetime. In Section~\ref{sect3}, we derive the unified expression for the flight times up to the 1PM order of relativistic massive and massless particles traveling in the field of the moving lens, and a discussion of the effect of the radial lens motion on the first-order Schwarzschild time delay of relativistic massive particles is then performed in the slow-motion limit. Section~\ref{sect4} is devoted to an analysis of the magnitude of the full radial velocity effect of the lens on the Schwarzschild delay and an evaluation of the possibility of its astronomical detection. A summary is given in Section~\ref{sect5}. Throughout this paper, the metric signature $(-,~+,~+,~+)$ and the geometrized units where $G=c=1$ are adopted. Greek indices run over $0,~1,~2$, and $3$ conventionally, and Latin indices run over $1,~2$, and $3$, unless indicated otherwise.

\begin{figure*}
\centering
\begin{minipage}[b]{16.5cm}
\includegraphics[width=16.5cm]{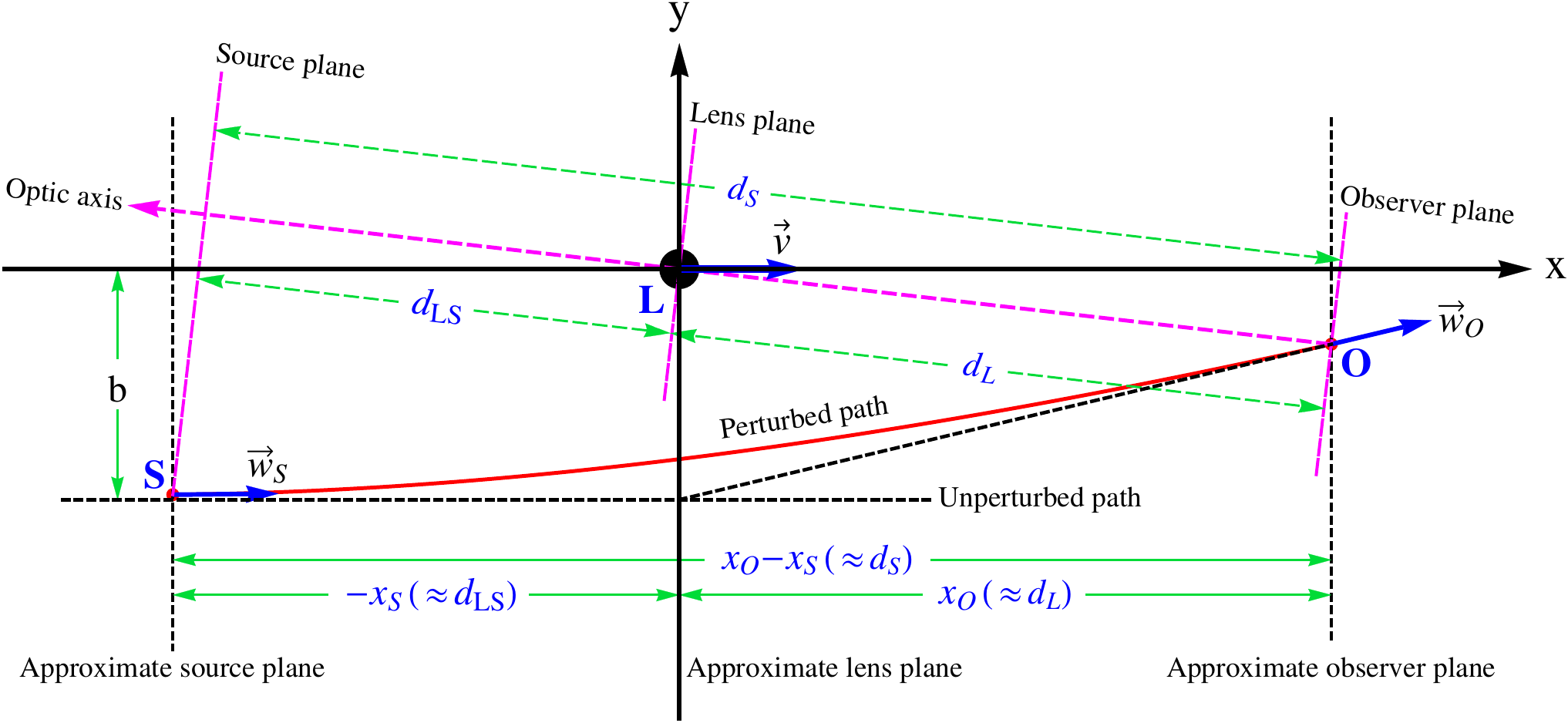}
\caption{Geometrical diagram for the propagation of a relativistic particle flying from the source $S$ to the observer $O$ in the equatorial plane (i.e., the $x$-$y$ plane) of a radially moving Schwarzschild black hole. The gravitational bending effect is exaggerated greatly to distinguish the perturbed path from the unperturbed one. }   \label{Figure1}
\end{minipage}
\end{figure*}
\section{Weak-field metric of a moving Schwarzschild black hole and equations of motion} \label{sect2}
Suppose that $\left\{\bm{e}_1,~\bm{e}_2,~\bm{e}_3\right\}$ is the orthonormal basis of a three-dimensional Cartesian coordinate system. We also assume that $X^\alpha\equiv\left(T,~X,~Y,~Z\right)$ and $x^\alpha\equiv\left(t,~x,~y,~z\right)$ denote the rest coordinate frame of a moving gravitational source and that of the observer of the background, respectively. Then in the observer's rest frame, the line element for the geometry of a moving Schwarzschild black hole with an arbitrary constant velocity $\bm{v}=v_1\bm{e}_1+v_2\bm{e}_2+v_3\bm{e}_3$ in the 1PM approximation reads~\cite{2014HLa}:
\begin{eqnarray}
&&\nn ds^2=-\left[1-\frac{2(1+v^2)\gamma^2M}{R}\right]\!dt^2-\frac{8\hspace*{0.8pt}v_i\gamma^2M}{R}dt\hspace*{1pt}dx^i ~~~ \\
&&\hspace*{27.5pt}+\left[\left(1+\frac{M}{R}\right)^2\delta_{ij}+\frac{4\hspace*{0.8pt}v_i\hspace*{0.6pt}v_j\gamma^2M}{R}\right]\!dx^idx^j~,    \label{metric}
\end{eqnarray}
where $v=|\bm{v}|=\sqrt{\Sigma_i v_i^2}$, $\gamma=\left(1-v^2\right)^{-\frac{1}{2}}$ is the Lorentz factor, $M$ denotes the rest mass of the black hole, and $R=\sqrt{X^2+Y^2+Z^2}$. Additionally, the coordinates of the comoving frame $\left(T,~X,~Y,~Z\right)$ of the black hole are related to those of the observer's rest frame $\left(t,~x,~y,~z\right)$ by the following Lorentz transformation:
\begin{eqnarray}
&&X^\alpha=\Lambda^\alpha_\beta x^\beta ~,    \label{LT}
\end{eqnarray}
with
\begin{eqnarray}
&&\Lambda^0_0=\gamma~,    \label{Lambda1} \\
&&\Lambda^i_0=\Lambda^0_i=-v_i\gamma~,    \label{Lambda2} \\
&&\Lambda^i_j=\delta_{ij}+\frac{(\gamma-1)v_iv_j}{v^2}~.    \label{Lambda3}
\end{eqnarray}

As the first of a series of works intended to probe the gravitational time delay of timelike signals induced by a moving gravitational system, we consider the leading-order gravitational time delay of relativistic massive particles propagating in the equatorial plane ($z=\frac{\partial}{\partial z}=0$) of a moving Schwarzschild black hole with a constant radial velocity $\bm{v}=v_1\bm{e}_1=v\bm{e}_1$ (by assuming $v_2=v_3=0$). For this purpose, we adopt the explicit equations of motion up to the 1PM order of a test particle traveling from $x\rightarrow -\infty$ with a relativistic initial velocity $\bm{w}=w\bm{e}_1$ ($0.05\lesssim w\leq1$~\cite{LTPNFKFZ2020} and $v<w$~\cite{WS2004}) parallel to the lens velocity $\bm{v}$ in the geometry of the radially moving gravitational body, and express them in the observer's rest frame as~\cite{HL2017b}:
\begin{eqnarray}
&&\nn\dot{t}=\frac{1}{w}\!-\!\frac{\gamma^2\!\left[\left(1\!+\!v^2\right)\!\left(\frac{v}{w^2}\!-\!\frac{2}{w}\right)\!+\!v\!\left(3\!-\!v^2\right)\right]}{1-\frac{v}{w}}\frac{M}{\sqrt{X^2\!+\!b^2}}  \\
&&\hspace*{17.6pt}+\,\mathcal{O}\left(M^2\right)~,~~~~    \label{ME1}  \\
&&\nn\dot{x}=1-\frac{\gamma^2\!\left[\left(1+v^2\right)\left(1-\frac{2v}{w}\right)-\frac{1-3v^2}{w^2}\right]}{1-\frac{v}{w}}\frac{M}{\sqrt{X^2+b^2}}  \\
&&\hspace*{17.6pt}+\,\mathcal{O}\left(M^2\right)~,~~~~    \label{ME2}  \\
&&\nn\dot{y}=\frac{\gamma\!\left[\left(1+v^2\right)\left(1+\frac{1}{w^2}\right)\!-\!\frac{4v}{w}\right]}{1-\frac{v}{w}}\frac{M}{b}\!\left(1\!+\!\frac{X}{\sqrt{X^2\!+\!b^2}}\right)  \\
&&\hspace*{17.6pt}+\,\mathcal{O}\left(M^2\right)~,~~~~    \label{ME3}
\end{eqnarray}
where $X=\gamma(x-vt)$, $b\,(\gg M)$ is the impact parameter, a dot denotes the derivative with respect to the trajectory parameter $\xi$ which is assumed to have the dimension of length, and the boundary conditions
$\dot{t}|_{\xi\rightarrow -\infty}=\dot{t}|_{x\rightarrow -\infty}=1/w$, $\dot{x}|_{\xi\rightarrow -\infty}=\dot{x}|_{x\rightarrow -\infty}=1$, $\dot{y}|_{\xi\rightarrow -\infty}=\dot{y}|_{x\rightarrow -\infty}=0$, and $y|_{\xi\rightarrow -\infty}=y|_{x\rightarrow -\infty}=-b$ have been used.

\section{Leading-order gravitational time delay of relativistic massive particles by a radially moving Schwarzschild lens} \label{sect3}
In this section, we first derive a new unified expression for the travel times of relativistic massive and massless particles propagating in the field of the radially moving Schwarzschild black hole within the 1PM framework, and then discuss the influence of the radial lens motion on the leading-order Schwarzschild delay of the massive particles in the slow-motion approximation.

Figure~\ref{Figure1} presents the corresponding geometrical diagram for the propagation of a relativistic particle, which is emitted by the source $S$ at the time $t=t_S$ and received by the observer $O$ at the time $t=t_O$, in the equatorial plane of the moving Schwarzschild lens $L$. The dashed horizontal line denotes the unperturbed trajectory of the particle with the mentioned initial velocity $\bm{w}\,(=w\bm{e}_1)$, while the red line stands for its perturbed propagation path due to the lens. In the observer's rest frame, the spatial coordinates of the source and the observer are denoted by $\left(x_S,~y_S,~0\right)$ and $\left(x_O,~y_O,~0\right)$, respectively, while their coordinates are denoted respectively by $\left(X_{S},~Y_{S},~0\right)$ and $\left(X_{O},~Y_{O},~0\right)$ in the comoving frame. Conventionally, the gravitational lens $L$, which is always situated at the origin of the comoving frame, is assumed to be located at the origin of the observer's rest frame when $t=0$. In the weak-field limit, both $S$ and $O$ are assumed to be situated in the asymptotically flat region, which means that they are at large but finite distances from the lens to ensure the convergence of the traveling time of the particle. We thus have $x_{S}\ll-b$, $x_{O}\gg b$, $y_{S}=-b\left[1+\mathcal{O}\left(M\right)\right]$~\cite{HL2016b}, $|\bm{w}_S|=w+\mathcal{O}\left(M\right)$, and $|\bm{w}_O|=w+\mathcal{O}\left(M\right)$, where $\bm{w}_S$ and $\bm{w}_O$ denote the particle's traveling velocities at points $S$ and $O$, respectively. The angular diameter distances of the source and the lens from the observer are denoted by $d_S$ and $d_L$, respectively, and $d_{LS}$ denotes the angular diameter distance of the source from the lens. Since the angle between the $x$-axis and the optic axis which joins the lens and the observer is actually tiny in general in our scenario, which leads to $d_{LS}\approx-x_{S}$, $d_{L}\approx x_{O}$, and $d_S\approx x_O-x_S$, we could regard the approximate source, lens, and observer planes as the exact ones (see the pink dashed parallel lines in Fig.~\ref{Figure1}), respectively, as done in \cite{HL2016b,HL2017b}.

\begin{widetext}
\subsection{The new unified formula for flight times of relativistic particles}  \label{sect3-1}
The unified expression for the flight times of the relativistic particles traveling in the equatorial plane of the radially moving Schwarzschild lens can be derived
according to Eqs.~\eqref{ME1} - \eqref{ME3} in the 1PM approximation. The substitution of Eq.~\eqref{ME2} into the integration of $\dot{t}$ shown in Eq.~\eqref{ME1} over $\xi$ yields the weak-field form of the coordinate time
\begin{eqnarray}
&&\nn t=\frac{x}{w}+\int\!\Bigg{\{}\!\left\{\frac{1}{w}-\frac{\gamma^2}{1-\frac{v}{w}}\!\left[\left(1+v^2\right)
\left(\frac{v}{w^2}-\frac{2}{w}\right)+v\left(3-v^2\right)\right]\!\frac{M}{\sqrt{X^2+b^2}}\right\}   \\
&&\hspace*{16.5pt} \times\left\{1+\frac{\gamma^2}{1-\frac{v}{w}}\!\left[\left(1+v^2\right)\!\left(1-\frac{2v}{w}\right)-\frac{1-3v^2}{w^2}\right]\!\frac{M}{\sqrt{X^2+b^2}}\right\}
-\frac{1}{w}\Bigg{\}}\,dx+C+\mathcal{O}\left(M^2\right)~,  \label{t1PM-Int}
\end{eqnarray}
where $C$ is an integral constant. By using the zeroth-order coordinate transformation $dx=\!\left[\gamma^{-1}\!\left(1\!-\!\frac{v}{w}\right)^{-1}+\mathcal{O}\left(M\right)\right]\!dX$ given in~\cite{HL2017b}, Eq.~\eqref{t1PM-Int} becomes
\begin{eqnarray}
t=\frac{x}{w}+\frac{(1-vw)\gamma\left[(3-v^2)w^2-1-4vw+3v^2\right]M}{w(w-v)^2}\ln\left(\sqrt{X^2+b^2}+X\right)+C+\mathcal{O}\left(M^2\right)~,    \label{t1PM}
\end{eqnarray}
which further leads to a unified formula for the travel times up to the 1PM order of relativistic massive $(0.05\lesssim w<1)$ and massless $(w=1)$ particles propagating from the source $S$ to the observer $O$ in the gravitational field of the moving lens
\begin{eqnarray}
\Delta t=\frac{x_O-x_S}{w}+\frac{(1-vw)\gamma\left[(3-v^2)w^2-1-4vw+3v^2\right]M}{w(w-v)^2} \ln\left(\frac{\sqrt{X_O^2+b^2}+X_O}{\sqrt{X_S^2+b^2}+X_S}\right)+\mathcal{O}\left(M^2\right)~,    \label{deltat-1PM-1}
\end{eqnarray}
\end{widetext}
where $X_S=\gamma(x_S-v\hspace*{0.7pt}t_S)$ and $X_O=\gamma(x_O-v\hspace*{0.7pt}t_O)$.

With respect to Eq.~\eqref{deltat-1PM-1}, four points are worth mentioning. First, we can see that the first term on the right-hand side of Eq.~\eqref{deltat-1PM-1} is the geometrical time for the particles traveling in a straight line when the lens is absent, while the second term there represents the expected compact expression for the leading-order gravitational time delays of the relativistic massive and massless particles induced by the moving lens. Second, the same result as Eq.~\eqref{deltat-1PM-1} can also be derived via assuming that the trajectory parameter has the dimension of time (see the Appendix~\ref{ApendA} for details). Third, it is interesting to find that in the case $w=1$, Eq.~\eqref{deltat-1PM-1} reduces to the result for the weak-field travel time of a light signal in the spacetime of the radially moving Schwarzschild lens achieved by a semi-iterative approach~\cite{HL2016b}
\begin{eqnarray}
&&\nn\left.\Delta t\right|_{w=1}=x_O\!-\!x_S+2\left(1\!-\!v\right)\gamma\,M\ln\!\left(\frac{\sqrt{X_O^2\!+\!b^2}\!+\!X_O}{\sqrt{X_S^2\!+\!b^2}\!+\!X_S}\right)  \\
&&\hspace*{1.57cm}+\,\mathcal{O}\left(M^2\right)~.   \label{deltat-1PM-light}
\end{eqnarray}
Note that the consistency between Eq.~\eqref{deltat-1PM-light} and the result obtained via the Li\'{e}nard-Wiechert representation~\cite{KS1999,KM2002} has been demonstrated in \cite{HL2016b}.

Fourth, a more convenient and practical form of Eq.~\eqref{deltat-1PM-1} can be expressed fully in terms of the quantities of the observer's rest frame. By assuming that the constant of integration in Eq.~\eqref{t1PM} is of 0PM order and that $C=0+\mathcal{O}(M^2)$ as an alternative choice, we have
\begin{eqnarray}
&&t_S=\frac{x_S}{w}+\mathcal{O}(M)~,        \label{tA} \\
&&t_O=\frac{x_O}{w}+\mathcal{O}(M)~,~~~~    \label{tB}
\end{eqnarray}
which give
\begin{eqnarray}
&&X_S=\gamma\left(1-\frac{v}{w}\right)x_S+\mathcal{O}(M)~,    \label{XA} \\
&&X_O=\gamma\left(1-\frac{v}{w}\right)x_O+\mathcal{O}(M)~.    \label{XB}
\end{eqnarray}
The substitution of Eqs.~\eqref{XA} - \eqref{XB} into Eq.~\eqref{deltat-1PM-1}, up to the 1PM order, then yields
\begin{widetext}
\begin{eqnarray}
\Delta t=\frac{x_O-x_S}{w}\!+\!\frac{\left(1\!-\!vw\right)\gamma\!\left[(3\!-\!v^2)w^2\!-\!1\!-\!4vw\!+\!3v^2\right]\!M}{w\left(w-v\right)^2}
\ln\!\!\left[\!\frac{\sqrt{\gamma^2\!\left(1\!-\!\frac{v}{w}\right)^2x_O^2\!+\!b^2}\!+\!\gamma\!\left(1\!-\!\frac{v}{w}\right)x_O}
{\sqrt{\gamma^2\!\left(1\!-\!\frac{v}{w}\right)^2x_S^2\!+\!b^2}+\gamma\!\left(1\!-\!\frac{v}{w}\right)x_S}\!\right]\!
\!+\!\mathcal{O}\left(M^2\right)~.~~~~~~~    \label{deltat-1PM-2}
\end{eqnarray}
With the consideration of $b\ll|x_S|$, $b\ll x_O$, and the astronomical notations $d_L\approx x_O$, $d_{LS}\approx -x_S$, and $d_S\approx x_O-x_S$, Eq.~\eqref{deltat-1PM-2} finally becomes
\begin{eqnarray}
\Delta t=\frac{d_S}{w}\!+\!\frac{\left(1\!-\!vw\right)\gamma\left[(3\!-\!v^2)w^2\!-\!1\!-\!4vw\!+\!3v^2\right]M}{w\left(w-v\right)^2}
\ln\!\left[4\gamma^2\left(1-\frac{v}{w}\right)^2\frac{d_{L}d_{LS}}{b^2}\right]+\mathcal{O}\left(M^2\right)~,~~~~~~    \label{deltat-1PM-3}
\end{eqnarray}
which, along with Eq.~\eqref{deltat-1PM-1}, serves as one of our main results.

\subsection{A typical astrophysical scenario: $|v|\ll 1$ and $w\rightarrow 1$} \label{sect3-2}
Considering that the slow-motion approximation $(|v|\ll 1)$ is valid for most astrophysical systems and is of astronomical interest, we first discuss the influence of the slow radial motion of the central body on the leading-order Schwarzschild delay of an ultrarelativistic massive particle $(w\rightarrow 1)$ such as a neutrino or an ultrahigh-energy cosmic-ray particle~\cite{BS2000,KO2011}.

Up to the first order in $v$, Eq.~\eqref{deltat-1PM-3} yields
\begin{eqnarray}
\Delta t=\frac{d_S}{w}+\left[\frac{3}{w}-\frac{1}{w^3}-\frac{v\left(2-3w^2+3w^4\right)}{w^4}\right]
\!M\ln\!\left(\frac{4d_{L}d_{LS}}{b^2}\right)+\frac{2v\left(1-3w^2\right)M}{w^4}+\mathcal{O}\left(M^2,~v^2\right)~,~~~~~~   \label{deltat-1PM-4}
\end{eqnarray}
which immediately gives the difference between the travel times of a relativistic massive particle $(0.05\lesssim w<1)$ and a light signal $(w=1)$
\begin{eqnarray}
\Delta(\Delta t)_1=\left(\frac{1}{w}\!-\!1\right)\!d_S+\!\left[\frac{3}{w}\!-\!\frac{1}{w^2}\!-\!2\!-\!\frac{v\left(2\!-\!3w^2\!+\!w^4\right)}{w^4}\right]\!M\ln\!\left[\frac{4d_{L}d_{LS}}{b^2}\right]
\!+\!\frac{2v\left(1\!-\!w^2\right)\!\left(1\!-\!2w^2\right)\!M}{w^4}+\mathcal{O}\left(M^2,~v^2\right)~,~~~~~~~   \label{DeltaDeltat1}
\end{eqnarray}
as well as the one between the travel times of two arbitrary relativistic massive particles whose initial velocities are respectively denoted by $w_1$ and $w_2$
\begin{eqnarray}
&&\nn\Delta(\Delta t)_2=\left(\frac{1}{w_2}-\frac{1}{w_1}\right)\!\Bigg{\{} d_S-\frac{2v\left(w_1+w_2\right)\left(3w_1^2w_2^2-w_1^2-w_2^2\right)M}{w_1^3w_2^3}+M\ln\!\left[\frac{4d_{L}d_{LS}}{b^2}\right]   \\
&&\hspace*{47pt}\times\,\frac{w_1^2w_2^2\left(3w_1w_2\!-\!1\right)\!-\!w_1w_2\left(w_1^2\!+\!w_2^2\right)+v\left(w_1\!+\!w_2\right)\!\left[3w_1^2w_2^2\!-\!2\left(w_1^2\!+\!w_2^2\right)\right]}{w_1^3w_2^3}\!\Bigg{\}}
+\mathcal{O}\left(M^2,~v^2\right)~,~~~~~~~~ \label{DeltaDeltat2}
\end{eqnarray}
in the spacetime of a moving Schwarzschild lens with a low radial velocity. For an ultrarelativistic massive particle $(w\rightarrow 1)$, we further perform a power series expansion of Eq.~\eqref{deltat-1PM-4} in a small parameter $\delta\equiv1-w~(\delta\ll1)$, which denotes the deviation degree of the particle's initial velocity from the speed of light, and have
\begin{eqnarray}
\Delta t=\left(1+\delta\right)d_S+2M\ln\!\left(\frac{4d_Ld_{LS}}{b^2}\right)-2v\left(1+\delta\right)M\!\left[2+\ln\!\left(\frac{4d_Ld_{LS}}{b^2}\right)\right]
+\mathcal{O}\left(M^2,~v^2,~\delta^2\right)~,~~~~~~   \label{deltat-1PM-5}
\end{eqnarray}
where the second- and higher-order terms in $\delta$ have been omitted.

According to Eq.~\eqref{deltat-1PM-5}, we find that the radial lens motion towards the observer speeds up the propagation of an ultrarelativistic massive particle traveling from the source to the observer by decreasing its flight time, when compared with the case in the Schwarzschild spacetime. On the contrary, if the lens gets away from the observer radially, the propagation process and the travel time of the massive particle under the same conditions will decelerate and increase, respectively. These conclusions are consistent with the ones revealed by the lightlike results ($\delta=0$) shown in the literature (see, e.g.,~\cite{HL2016b,KS1999}). Moreover, the $v$-dependent term on the right-hand side of Eq.~\eqref{deltat-1PM-5} can be conventionally expressed as
\begin{eqnarray}
\Delta t_M=-\,0.17\left(\frac{v}{0.00073}\right)\!\left(\frac{1+\delta}{1.000001}\right)\!\left(\frac{M}{10^6M_{\bullet}}\right)\!
\left\{1+0.042\ln\!\left[\left(\frac{d_L}{8.2\,\text{kpc}}\right)\!\left(\frac{d_{LS}}{0.01\,\text{kpc}}\right)\!\left(\frac{b}{10^{-5}\,\text{kpc}}\right)^{-2}\right]\right\}\,\text{s}~.~~~~~~~   \label{deltat-v}
\end{eqnarray}
Additionally, Eq.~\eqref{deltat-1PM-5} gives the difference between the weak-field gravitational time delays of an ultrarelativistic massive particle and a light signal caused by the slowly moving lens
\begin{eqnarray}
\Delta(\Delta t)_{1M}\simeq -\,2\hspace*{1pt}v\hspace*{1pt}\delta\hspace*{1pt}M\!\left[2+\ln\!\left(\frac{4d_Ld_{LS}}{b^2}\right)\right]~.~~~~~~   \label{DeltaDeltat1b}
\end{eqnarray}
For two ultrarelativistic massive particles with initial velocities $w_{1}\equiv 1-\delta_1$ and $w_{2}\equiv 1-\delta_2$, respectively, the difference between their weak-field gravitational delays due to the same lens can also be obtained from Eq.~\eqref{deltat-1PM-5} as
\begin{eqnarray}
\Delta(\Delta t)_{2M}\simeq-\,2\hspace*{1pt}v\left(\delta_2-\delta_1\right)M\!\left[2+\ln\!\left(\frac{4d_Ld_{LS}}{b^2}\right)\right]~.~~~~~~   \label{DeltaDeltat2b}
\end{eqnarray}

\begin{table}
\begin{minipage}[t]{\textwidth}
\begin{tabular}{cccccccc}
\toprule[1.3px]
     $v\:\backslash\:w$~~~~     &            $ 0.05 $           &              $0.1$            &             $0.5$             &             $0.9$             &             $0.95$            &            $1$             \\
\midrule[0.5pt]   \vspace*{-6pt}  \\
                     $-$0.9~~~~ &              20.8             &               2.6             &       4.2$\times10^{-2}$      &       2.1$\times10^{-2}$      &       2.0$\times10^{-2}$      &    ~~~~1.9$\times10^{-2}$  \\
                     $-$0.5~~~~ &              20.5             &               2.5             &       1.6$\times10^{-2}$      &       4.5$\times10^{-3}$      &       4.3$\times10^{-3}$      &    ~~~~4.1$\times10^{-3}$  \\
                     $-$0.1~~~~ &              18.2             &               1.9             &       4.8$\times10^{-3}$      &       6.7$\times10^{-4}$      &       6.1$\times10^{-4}$      &    ~~~~5.8$\times10^{-4}$  \\
                   $-$0.001~~~~ &               0.8             &       4.9$\times10^{-2}$      &       5.9$\times10^{-5}$      &       6.4$\times10^{-6}$      &       5.8$\times10^{-6}$      &    ~~~~5.5$\times10^{-6}$  \\
                $-$0.000001~~~~ &       8.0$\times10^{-4}$      &       5.0$\times10^{-5}$      &       5.9$\times10^{-8}$      &       6.4$\times10^{-9}$      &       5.8$\times10^{-9}$      &    ~~~~5.5$\times10^{-9}$  \\
                   0.000001~~~~ &  ~~~~$-$8.0$\times10^{-4}$~~~~& ~~~~$-$5.0$\times10^{-5}$~~~~ & ~~~~$-$5.9$\times10^{-8}$~~~~ & ~~~~$-$6.4$\times10^{-9}$~~~~ & ~~~~$-$5.8$\times10^{-9}$~~~~ &  ~~~~$-$5.5$\times10^{-9}$ \\
                      0.001~~~~ &             $-$0.8            &     $-$5.0$\times10^{-2}$     & ~~~~$-$5.9$\times10^{-5}$~~~~ & ~~~~$-$6.4$\times10^{-6}$~~~~ & ~~~~$-$5.8$\times10^{-6}$~~~~ &  ~~~~$-$5.5$\times10^{-6}$ \\
                        0.1~~~~ &            $\star$            &             $\star$           & ~~~~$-$7.7$\times10^{-3}$~~~~ & ~~~~$-$6.2$\times10^{-4}$~~~~ & ~~~~$-$5.6$\times10^{-4}$~~~~ &  ~~~~$-$5.2$\times10^{-4}$ \\
                        0.5~~~~ &            $\star$            &             $\star$           &            $\star$            & ~~~~$-$3.1$\times10^{-3}$~~~~ & ~~~~$-$2.5$\times10^{-3}$~~~~ &  ~~~~$-$2.3$\times10^{-3}$ \\
                        0.9~~~~ &            $\star$            &             $\star$           &            $\star$            &            $\star$            & ~~~~$-$9.5$\times10^{-3}$~~~~ &  ~~~~$-$4.1$\times10^{-3}$ \\
\bottomrule[1.3px]
\end{tabular} \par  \vspace*{3pt}
\end{minipage}
\caption{The values (in units of s) of $\delta\Delta t$ for the case of the stellar-mass black hole Cyg X-1. Here and thereafter, the special case of light $(w=1)$ is listed for comparison, a star ``$\star$" denotes a value which is not under consideration due to $v<w$, and our attention is concentrated on the absolute values of $\delta\Delta t$. Additionally, we present the cases with high values of $|v|$ mainly for illustration, since the possibility that a massive black hole moves at a relativistic velocity is very small or even non-existent. }   \label{Table1}
\end{table}

\begin{table}
\begin{minipage}[t]{\textwidth}
\begin{tabular}{cccccccc}
\toprule[1.3px]
$v\:\backslash\:w$~~~~  &            $ 0.05 $           &              $0.1$            &             $0.5$             &             $0.9$             &             $0.95$            &            $1$             \\
\midrule[0.5pt]   \vspace*{-6pt}  \\
             $-$0.9~~~~ &        4.5$\times10^{4}$      &        5.7$\times10^{3}$      &              90.2             &              44.6             &              42.9             &        ~~~~41.5            \\
             $-$0.5~~~~ &        4.4$\times10^{4}$      &        5.4$\times10^{3}$      &              35.5             &               9.7             &               9.2             &         ~~~~8.8            \\
             $-$0.1~~~~ &        3.9$\times10^{4}$      &        4.1$\times10^{3}$      &              10.4             &               1.4             &               1.3             &         ~~~~1.2            \\
           $-$0.001~~~~ &        1.7$\times10^{3}$      &        1.1$\times10^{2}$      &               0.1             &       1.4$\times10^{-2}$      &       1.3$\times10^{-2}$      &    ~~~~1.2$\times10^{-2}$  \\
        $-$0.000001~~~~ &               1.7             &                0.1            &       1.3$\times10^{-4}$      &       1.4$\times10^{-5}$      &       1.3$\times10^{-5}$      &    ~~~~1.2$\times10^{-5}$  \\
           0.000001~~~~ &            $-$1.7             &              $-$0.1           & ~~~~$-$1.3$\times10^{-4}$~~~~ & ~~~~$-$1.4$\times10^{-5}$~~~~ & ~~~~$-$1.3$\times10^{-5}$~~~~ &  ~~~~$-$1.2$\times10^{-5}$ \\
              0.001~~~~ & ~~~~$-$1.8$\times10^{3}$~~~~  & ~~~~$-$1.1$\times10^{2}$~~~~ &            $-$0.1             & ~~~~$-$1.4$\times10^{-2}$~~~~ & ~~~~$-$1.3$\times10^{-2}$~~~~ &  ~~~~$-$1.2$\times10^{-2}$  \\
                0.1~~~~ &            $\star$            &             $\star$           &           $-$16.7             &             $-$1.3            &             $-$1.2            &      ~~~~$-$1.1            \\
                0.5~~~~ &            $\star$            &             $\star$           &            $\star$            &             $-$6.8            &             $-$5.5            &      ~~~~$-$4.9            \\
                0.9~~~~ &            $\star$            &             $\star$           &            $\star$            &            $\star$            &            $-$20.5            &      ~~~~$-$8.8            \\
\bottomrule[1.3px]
\end{tabular} \par  \vspace*{3pt}
\end{minipage}
\caption{The values (in units of s) of $\delta\Delta t$ for the case of the intermediate-mass black hole NGC 3319$^\ast$. }   \label{Table2}
\end{table}

\begin{table}
\begin{minipage}[b]{\textwidth}
\begin{tabular}{cccccccc}
\toprule[1.3px]
$v\:\backslash\:w$~~~~  &            $ 0.05 $           &              $0.1$            &             $0.5$             &             $0.9$             &             $0.95$            &            $1$              \\
\midrule[0.5pt]   \vspace*{-6pt}  \\
             $-$0.9~~~~ &        4.1$\times10^{6}$      &        5.2$\times10^{5}$      &       8.5$\times10^{3}$       &       4.2$\times10^{3}$       &       4.1$\times10^{3}$       &  ~~~~3.9$\times10^{3}$      \\
             $-$0.5~~~~ &        4.0$\times10^{6}$      &        4.9$\times10^{5}$      &       3.3$\times10^{3}$       &       9.1$\times10^{2}$       &       8.6$\times10^{2}$       &  ~~~~8.2$\times10^{2}$      \\
             $-$0.1~~~~ &        3.6$\times10^{6}$      &        3.7$\times10^{5}$      &       9.4$\times10^{2}$       &       1.3$\times10^{2}$       &       1.2$\times10^{2}$       &  ~~~~1.2$\times10^{2}$      \\
           $-$0.001~~~~ &        1.5$\times10^{5}$      &        9.4$\times10^{3}$      &              11.4             &               1.3             &               1.2             &        ~~~~1.1              \\
        $-$0.000001~~~~ &        1.5$\times10^{2}$      &                9.6            &       1.1$\times10^{-2}$      &       1.3$\times10^{-3}$      &       1.2$\times10^{-3}$      &  ~~~~1.1$\times10^{-3}$     \\
           0.000001~~~~ & ~~~~$-$1.5$\times10^{2}$~~~~  &              $-$9.6           & ~~~~$-$1.1$\times10^{-2}$~~~~ & ~~~~$-$1.3$\times10^{-3}$~~~~ & ~~~~$-$1.2$\times10^{-3}$~~~~ &  ~~~~$-$1.1$\times10^{-3}$  \\
              0.001~~~~ & ~~~~$-$1.6$\times10^{5}$~~~~  &  ~~~~$-$9.7$\times10^{3}$~~~~ &           $-$11.5             &             $-$1.3            &             $-$1.2            &     ~~~~$-$1.1              \\
                0.1~~~~ &            $\star$            &             $\star$           &      $-$1.5$\times10^{3}$     &  ~~~~$-$1.2$\times10^{2}$~~~~ &  ~~~~$-$1.1$\times10^{2}$~~~~ &  ~~~~$-$1.0$\times10^{2}$   \\
                0.5~~~~ &            $\star$            &             $\star$           &            $\star$            &  ~~~~$-$6.2$\times10^{2}$~~~~ &  ~~~~$-$5.0$\times10^{2}$~~~~ &  ~~~~$-$4.5$\times10^{2}$   \\
                0.9~~~~ &            $\star$            &             $\star$           &            $\star$            &            $\star$            &  ~~~~$-$1.8$\times10^{3}$~~~~ &  ~~~~$-$8.1$\times10^{2}$   \\
\bottomrule[1.3px]
\end{tabular} \par  \vspace*{3pt}
\end{minipage}
\caption{The values (in units of s) of $\delta\Delta t$ for the case of the supermassive black hole Sgr A$^\ast$. }   \label{Table3}
\end{table}

\end{widetext}

\section{Radial velocity effect on leading-order Schwarzschild delay of massive particles} \label{sect4}
Since the influence of the radial motion of the lens on the first-order gravitational time delay of light signals has been considered in the literature~\cite{KS1999,Sereno2002,KM2002,Frittelli2003,Sereno2005,KF2007,BAL2008,Kopeikin2009,HBL2014,SH2014,HL2016b}, we then discuss the full radial velocity effect on the leading-order Schwarzschild time delay of the relativistic massive particles, the expression of which is given by Eq.~\eqref{deltat-1PM-3} as
\begin{eqnarray}
&&\nn\delta\Delta t=M\Bigg{\{}\frac{\left(1-vw\right)\gamma\left[\left(3-v^2\right)w^2-1-4vw+3v^2\right]}{w\left(w-v\right)^2}   \\
&&\nn\hspace*{11pt}\times\ln\!\left[4\gamma^2\!\left(1\!-\!\frac{v}{w}\right)^2\frac{d_{L}d_{LS}}{b^2}\right]\!-\!\frac{3w^2\!-\!1}{w^3}\ln\!\left[\frac{4d_{L}d_{LS}}{b^2}\right]\!\!\Bigg{\}}   \\
&&\hspace*{11pt}+\,\mathcal{O}\left(M^2\right)~.~~~~~   \label{VE-1}
\end{eqnarray}
Note that $\delta\Delta t$ vanishes in the case of no lens motion $(v=0)$.

Up to now, it is widely believed that there are three populations of black holes in the universe by their masses, namely stellar-mass ($M\sim 3-20M_{\odot}$), supermassive ($M\sim 10^6-10^{10}M_{\odot}$), and intermediate-mass ($M\sim 10^2-10^5M_{\odot}$) black holes~\cite{GSH2020,MC2004,CJ2014,MR2001,NM2013}, with $M_{\odot}$ ($\simeq4.92\mu $s) being the rest mass of the sun. In order to estimate the order of magnitude of $\delta\Delta t$,
we model the stellar-mass black hole in Cygnus X-1 (i.e., Cyg X-1 with $M\simeq14.8M_{\odot}$~\cite{OMARRNG2011} and $d_{L}\simeq1.9$ kpc~\cite{RMNGRO2011,Gou2011}),
the intermediate-mass black hole candidate in the galaxy NGC 3319 (namely, NGC 3319$^\ast$ with $M\simeq3.1\times10^{4}M_{\odot}$~\cite{DG2021,JWZ2018} and $d_{L}\simeq14.3$ Mpc~\cite{Sakai1999,JWZ2018}),
and the Galactic supermassive black hole (i.e., Sgr A$^\ast$ with $M\simeq4.2\times10^{6}M_{\odot}$~\cite{BG2016,Parsa2017} and $d_{L}\simeq8.2$ kpc~\cite{BG2016}) as the lens, respectively. With the consideration that typical values of the angular diameter distance of the light source from the lens are within the domain $0.001\,\text{kpc}\lesssim d_{LS}\lesssim  0.1\,\text{kpc}$~\cite{KP2005}, here we may preset $d_{LS}=0.01\,\text{kpc}$ for these three black-hole lens systems as an example. Moreover, we follow the idea of~\cite{IRDMA2008} and assume the impact parameter $b$ to be the natural Einstein ring radius $R_E$ of light signals of a black hole approximately, which yields $b\simeq5.3\times10^{-9}\,$kpc, $2.4\times10^{-7}\,$kpc, and $2.8\times10^{-6}\,$kpc for Cyg X-1, NGC 3319$^\ast$, and Sgr A$^\ast$, respectively. The ranges of $w$ and $v$ have been given above, namely $w\in\left[0.05,~1\right]$~\cite{LTPNFKFZ2020} and $v\in\left(-1,~w\right)$~\cite{WS2004}, where $w=1$ denotes the special case of light. Tables~\ref{Table1} - \ref{Table3} present the values of $\delta\Delta t$ for the scenarios of Cyg X-1, NGC 3319$^\ast$, and Sgr A$^\ast$, respectively, according to which three points should be mentioned. First, for a massive particle traveling with a low relativistic initial velocity, it is found that the absolute values of $\delta\Delta t$ for black holes with different rest masses under the above assumption of $b=R_E$ can be evidently larger than the time resolutions of some of current major detectors or instruments, whether or not the radial motions of those lenses take are relativistic. For example, if $w=0.1$ and a very low radial lens velocity $v=1.0\times10^{-6}\,(\sim 0.3\,\text{km/s})$ are given, $|\delta\Delta t|$ for Cyg X-1, NGC 3319$^\ast$, and Sgr A$^\ast$ will have values 49.5$\mu$s, 0.1s, and 9.6s, respectively, all of which are at least four orders of magnitude larger than the time resolution ($\sim$\,2ns) of the LHAASO project~\cite{Cao2024,Ma2022}. Second, for a massive particle (e.g., a high-energy cosmic-ray particle) with a moderate relativistic initial velocity, the values of the velocity effect $\delta\Delta t$ under the same conditions can also be larger than the time resolutions of those detectors in almost all of the relativistic cases, as well as in most of the nonrelativistic cases, of the radial lens motions. $|\delta\Delta t|$ for the cases of Cyg X-1, NGC 3319$^\ast$, and Sgr A$^\ast$ is respectively about 3, 6.9$\times10^3$, and 6.4$\times10^5$ times larger than the time resolution of the LHAASO, when $w=0.9$ and $v=1.0\times10^{-6}$ are assumed. A final point is that for a massive particle with a highly relativistic initial velocity, the velocity effect on the gravitational delay can still be noticeably larger than the time resolutions of current major projects
(e.g., the IceCube neutrino detector~\cite{IceCube2006,Abbasi2010,HK2010} and the ANTARES neutrino telescope~\cite{Feinstein2003,Ageron2007,Brunner2011}) for most of the radial lens motions, although its value for the case of $v$, $b$, and the lens system being given decreases with the increase of $w$. For instance, even for an ordinary massive neutrino with an ultrarelativistic initial velocity $w=1-1.0\times10^{-6}$~\cite{Adamson2015,Adam2012} serving as the test particle, $|\delta\Delta t|$ for the black hole systems Cyg X-1, NGC 3319$^\ast$, and Sgr A$^\ast$ under the same conditions will be 3 to 9 orders of magnitude larger than the time resolution of the IceCube neutrino detector, provided that $v=0.001$ is given. Hence, similar to the case of light signals as test particles, the results given in Tabs.~\ref{Table1} - \ref{Table3} indicate a large possibility to measure the radial velocity effect on the leading-order Schwarzschild time delay of relativistic massive particles with the current (or near future) capability of some major projects.

\section{Summary} \label{sect5}
In summary, the leading-order gravitational time delay of relativistic massive particles induced by a radially moving Schwarzschild black hole has been probed in this work. We have achieved a new unified formula for the flight times up to the first post-Minkowskian order of relativistic massive and massless particles traveling in the field of the moving lens firstly. The expression of the difference between the flight times of a relativistic massive particle and a light signal in this spacetime, along with that of two relativistic massive particles, has thus been obtained in the slow-motion approximation. We have also discussed the effect of the radial lens motion on the leading-order Schwarzschild delay of the relativistic massive particles. Compared with the case in the Schwarzschild spacetime, it is found that the radial lens motion towards the observer speeds up the propagation of an ultrarelativistic massive particle traveling from the source to the observer and thus decreases its flight time in the weak-field and slow-motion limit. In contrast, provided that the lens moves away from the observer radially, the propagation process and the travel time of the massive particle under the same conditions will decelerate and increase, respectively. Moreover, the magnitude of the full radial velocity effect of the lens and the possibility of its astronomical detection have been analyzed in detail via modeling three representative black hole systems as the lens respectively. We find that there is a large possibility to detect the radial velocity effect on the leading-order Schwarzschild delay of relativistic massive particles with the capability of current (or near future) technologies, whether or not the lens motion is relativistic.

\begin{acknowledgments}
G.H. is funded partially by the National Natural Science Foundation of China (Grant No. 12205139). W.L. is funded in part by the National Natural Science Foundation of China (Grant No. 12475057). Y.X. is funded by the National Natural Science Foundation of China (Grant Nos. 12273116 and 62394351), the science research grants from the China Manned Space Project (Grant Nos. CMS-CSST-2021-A12 and CMS-CSST-2021-B10) and the Opening Project of National Key Laboratory of Aerospace Flight Dynamics of China (Grant No. KGJ6142210220201).
\end{acknowledgments}

\appendix

\begin{widetext}

\section{An alternative derivation of Eq.\,(11)} \label{ApendA}
In the computation of Eq.~\eqref{deltat-1PM-1} above, the trajectory parameter $\xi$ has been assumed to take the dimension of length. Actually, we can also adopt the assumption that $\xi$ has the dimension of time, under which the explicit geodesic equations in the observer's rest frame take the following form~\cite{HL2017b}:
\begin{eqnarray}
&&\dot{t}=1-\frac{\gamma^2\hspace*{-0.6pt}\left[\left(1+v^2\right)\left(\frac{v}{w}-2\right)+vw\left(3-v^2\right)\right]}{1-\frac{v}{w}}\frac{M}{\sqrt{X^2+b^2}}+\mathcal{O}\left(M^2\right)~,~~~~    \label{ME1-A}  \\
&&\dot{x}=w+\frac{w\gamma^2}{1\!-\!\frac{v}{w}}\!\left[\left(1\!+\!v^2\right)\left(\frac{2v}{w}\!-\!1\right)\!+\!\frac{1\!-\!3v^2}{w^2}\right]\!\frac{M}{\sqrt{X^2+b^2}}+\mathcal{O}\left(M^2\right)~,~~~~    \label{ME2-A}  \\
&&\dot{y}=\frac{\gamma\left[\left(1+v^2\right)\left(1+w^2\right)-4vw\right]}{w-v}\frac{M}{b}\left(1+\frac{X}{\sqrt{X^2+b^2}}\right)+\mathcal{O}\left(M^2\right)~.~~~~    \label{ME3-A}
\end{eqnarray}
Now we substitute the first-order parameter transformation obtained from Eq.~\eqref{ME2-A} into the integration of Eq.~\eqref{ME1-A} over $\xi$, and have
\begin{eqnarray}
&&\nn t=\frac{x}{w}+\frac{1}{w}\int\!\Bigg{\{}\!\left\{1-\frac{\gamma^2}{1-\frac{v}{w}}\left[\left(1+v^2\right)\left(\frac{v}{w}-2\right)+vw\left(3-v^2\right)\right]\frac{M}{\sqrt{X^2+b^2}}\right\} \\
&&\hspace*{17pt}\times\left\{1-\frac{\gamma^2}{1-\frac{v}{w}}\left[\left(1+v^2\right)\!\left(\frac{2v}{w}-1\right)+\frac{1-3v^2}{w^2}\right]\frac{M}{\sqrt{X^2+b^2}}\right\}-1\Bigg{\}}\,dx+C
+\mathcal{O}\left(M^2\right)~. \label{t1PM-Int-A}
\end{eqnarray}
Similarly, by using the 0PM coordinate transformation above, we can achieve the same results from Eq.~\eqref{t1PM-Int-A} as these presented in Eqs.~\eqref{t1PM} - \eqref{deltat-1PM-1}.
\end{widetext}

\end{document}